\documentclass[twocolumn,showpacs,nofootinbib,preprintnumbers,amsmath,amssymb,showkeys]{revtex4}

\usepackage{epsfig}
\usepackage{dsfont}

\usepackage{graphicx}
\usepackage{dcolumn}
\usepackage{bm}

\newcommand{\beq}{\begin{equation}}
\newcommand{\eeq}{\end{equation}}
\newcommand{\bea}{\begin{eqnarray}}
\newcommand{\eea}{\end{eqnarray}}
\newcommand{\nn}{\nonumber \\}

\newcommand\eqn[1]{(\ref{#1})}      
\newcommand\Eqn[1]{Eq.~(\ref{#1})}  

\newcommand{\C}{\mathcal{C}}

\newcommand{\bx}{{\bf x}}
\newcommand{\by}{{\bf y}}
\newcommand{\bs}{{\bf s}}
\newcommand{\bp}{{\bf p}}
\newcommand{\bk}{{\bf k}}
\newcommand{\bq}{{\bf q}}
\newcommand{\be}{{\bf e}}

\newcommand{\bl}{{\bf l}}

\newcommand{\bX}{{\bf X}}
\newcommand{\bK}{{\bf K}}
\newcommand{\bS}{{\bf S}}
\newcommand{\bQ}{{\bf Q}}

\newcommand{\bL}{{\bf L}}

\begin{document}


\title{Physical momentum representation of scalar field correlators \\ in 
de Sitter space}

\author{R. Parentani}
 \email{parentani@th.u-psud.fr}
\affiliation{%
 Laboratoire de Physique Th\'eorique (LPT), CNRS UMR 8627, B\^at. 210, Universit\'e Paris - Sud 11, 91405 Orsay Cedex, France
}%
\author{J. Serreau}%
 \email{serreau@apc.univ-paris7.fr}
\affiliation{%
 Astro-Particule et Cosmologie (APC), CNRS UMR 7164, Universit\'e Paris 7 - Denis Diderot\\ 10, rue Alice Domon et L\'eonie Duquet, 75205 Paris Cedex 13, France
}%

\date{\today}

\begin{abstract}

We propose a new approach to compute correlators of quantum fields in de Sitter space. It is based on nonequilibrium field theory techniques, and exploits de Sitter symmetries so as to partially reduce the number of independent variables of $n$-point functions in a manner that preserves the usefulness of a momentum representation, e.g., for writing spatial convolution integrals as simple products. In this representation, the two-point function of a scalar field only depends on two physical momenta, and the corresponding Schwinger-Dyson evolution equations take the form of momentum flow equations. Moreover, standard diagrammatic rules can be entirely formulated in this representation. The method is suitable for analytical approximations as well as numerical implementations. In forthcoming publications, we apply it to resum infrared logarithmic terms
appearing in the perturbative calculation of vertex and correlation functions.
 \end{abstract}

\pacs{11.10.-z, 04.62.+v}
\keywords{Quantum field theory, de Sitter space}
\maketitle


\section{Introduction}
\label{sec:intro}

The study of quantum fields in de Sitter space is a topic of timely interest. Although the issue of computing radiative corrections in curved spaces in general and in de Sitter space in particular is a rather old topic (see e.g. \cite{Birrell:1982ix}), it has seen a renewed interest in the last decade with strong motivations from recent cosmological observations. In particular, the impressive success of the inflationary paradigm in the early Universe \cite{Peiris:2003ff,Parentani:2004ta} and the observation of the recent acceleration of the Universe \cite{Perlmutter:1998np} motivate one to better understand quantum field theory (QFT) in expanding space-times. 

A fundamental question is the so-called trans-Planckian issue  \cite{Jacobson:1999zk}, 
i.e., the issue of an effective decoupling between high- and low-energy physics, which is at the root of the concept of effective QFT. If decoupling is rather well understood in flat space-time \cite{Weinberg:1996kw,Delamotte:2007pf}, the situation is much less clear in expanding universes, where gravitational redshift induces a kinematical correlation between infrared (IR) and (arbitrarily high) ultraviolet (UV) modes, for which an effective description in terms of a fixed background geometry may not be appropriate.
Clear light is shed on this issue when considering a lattice formulation of QFT in expanding universes \cite{Weiss:1985vw,Boyanovsky:1996rw,Baacke:1997rs,Jacobson:1999zk}. When working with a fixed number of lattice points, their density decreases as the universe expands, thereby limiting the number of e-foldings
of the simulation during which some useful information can be extracted\footnote{In addition, when working with a comoving lattice, the bare parameters of the theory must depend on the cosmological time in order to keep the renormalized parameters constant at a given physical scale \cite{Boyanovsky:1996rw,Serreau:2011fu}.};
see, e.g., \cite{Tranberg:2008ae}.
To avoid this dilution, one is led to consider a lattice where the number of sites increases as expansion proceeds. 
It is then a practical question how to initialize these incoming degrees of freedom \cite{Boyanovsky:1996rw} and how to couple them with preexistent configurations.

Besides this issue, one should also consider nonperturbative techniques. Radiative corrections in cosmological Friedmann-Robertson-Walker (FRW) space-times have been addressed in a variety of field theories \cite{Boyanovsky:1996rw,Prokopec:2002jn,Weinberg:2005vy,Anderson:2005hi}, mainly based on the perturbative loop expansion \cite{Birrell:1982ix}. Then loop diagrams typically exhibit secular terms, which grow as powers of the number of e-folds, and which can turn into severe infrared divergences when the field is light in units of the Hubble rate \cite{Starobinsky:1994bd,Weinberg:2005vy,Tsamis:2005hd}. Spurious secular terms are typical of perturbative approaches for nonequilibrium (time dependent) problems in QFT
formulated in flat space-time \cite{Berges:2004vw}. They prevent the study of late time evolution and must be resummed to get meaningful results. Similarly IR divergences naturally arise in situations with light bosonic degrees of freedom, such as scalar or gauge fields at high temperatures \cite{Blaizot:2003tw} or near a second order phase transition \cite{Delamotte:2007pf}. They usually signal a deficiency of the perturbative approach\footnote{Note, however, that in the context of inflationary cosmology, the authors of Refs. \cite{Urakawa:2009my} have argued that IR divergences do not affect gauge-invariant observables; see also \cite{Senatore:2009cf}.}. This calls for resummation techniques and/or nonperturbative methods possibly involving numerical techniques.

A number of methods have been developed over the years to deal with these issues, such as renormalization group \cite{Delamotte:2007pf,Boyanovsky:1998aa} or two-particle-irreducible (2PI) \cite{Blaizot:2003tw,Calzetta:1986cq,Berges:2004vw} techniques. Since the cosmological context is a nonequilibrium setup, the most appropriate tools are those of nonequilibrium QFT. Techniques such as the dynamical renormalization group \cite{Burgess:2009bs}, or the 2PI formalism \cite{Calzetta:1986ey,Ramsey:1997qc,Tranberg:2008ae,Garbrecht:2011gu} can be formulated for expanding space-times. Introducing conformal time and comoving spatial coordinates as well as conformally rescaled fields, the relevant equations for an interacting scalar field theory for instance actually very much resemble their Minkowski counterparts. The expansion is only manifest in an additional time-dependent mass term involving both the expansion rate and acceleration, see e.g. \cite{Tranberg:2008ae}. This {\it apparently} allows one to use the (numerical) tools developed for nonequilibrium QFT in flat space-time directly in this context. However, as discussed above, the gravitational redshift actually limits the numerical simulations to a low number of e-folds \cite{Tranberg:2008ae}. Therefore one must look for approaches that avoid this limitation.

Because of its larger degree of symmetry and of its relevance to inflationary cosmology, de Sitter space-time has been much investigated \cite{Prokopec:2002jn,Brunier:2004sb,Boyanovsky:2005px,Sloth:2006az,Seery:2007we,vanderMeulen:2007ah,Senatore:2009cf,Marolf:2010nz,Hollands:2010pr,Higuchi:2010xt,Boyanovsky:2012nd}. In this context, a very efficient effective description for the nonperturbative dynamics of IR modes has been devised, the so-called stochastic approach \cite{Starobinsky:1994bd}, which has been shown to actually resum the leading IR logarithms of perturbation theory to all orders \cite{Tsamis:2005hd}. It is, however, desirable to go beyond this effective description for various reasons, e.g. for computing corrections to the stochastic approach, or to address specific issues outside its domain of applicability, such as e.g. decoupling, which requires a dynamical description of both IR and UV modes. An interesting proposal to systematically include perturbative corrections to the results of the stochastic approach in the context of Euclidean de Sitter space has been put forward in \cite{Rajaraman:2010xd,Beneke:2012kn}. 
For scalar field theories, the large-$N$ \cite{Riotto:2008mv,Serreau:2011fu}, or Hartree \cite{Prokopec:2011ms,Arai:2011dd} approximations provide nonperturbative approaches capable of taking into account the full coupled dynamics of IR and UV modes. These essentially amount to a local mass resummation and are well suited to describe dynamical mass generation in de Sitter. Going beyond such mean-field-like descriptions typically requires one to treat nonlocal integro-differential equations \cite{Tranberg:2008ae}. Standard nonequilibrium techniques to deal with the latter suffer the same drawbacks described above for generic FRW geometries.

In the de Sitter geometry, one may hope to use the large symmetry group to further simplify the formulation and overcome the problems mentioned above. For instance, a two-point function in comoving momentum space in a FRW geometry with flat spatial sections depends on three variables: two (conformal) times and the modulus of a comoving momentum. One thus effectively deals with a ($1+1$)-dimensional problem \cite{Tranberg:2008ae}. In de Sitter space, two-point correlators in real space only depend on one variable, the de Sitter invariant distance. However, the equations of motion for correlators, which typically involve nonlocal convolution integrals are difficult to formulate in real space in a way suitable for both analytical and numerical calculations. Instead these are conveniently formulated in Fourier space where, however, the full de Sitter symmetry is not transparent and difficult to exploit.

Here, we propose an intermediate approach where we exploit only partially the full de Sitter group. Our approach is a momentum representation in that convolutions and loop integrals keep a simple form, suitable for numerical implementations as well as for simplified analytic treatments. It generalizes the so-called $p$-representation introduced in \cite{Busch:2012ne}, and exploits the fact that the expanding de Sitter space-time can be equivalently seen as time dependent and spatially homogeneous, or as stationary but spatially inhomogeneous. The subgroup which combines these two symmetries implies that the two-point correlation function (or the vertex function) of a scalar field can be expressed in terms of only two physical momenta. We then show that the relevant Schwinger-Dyson (SD) equations for two-point functions take a particularly simple form in terms of these variables: the time evolution equation formally becomes flow equations in physical momentum space. In addition, this effectively reduces to a ($0+1$)--dimensional problem.

We thus see that this approach combines mathematical simplifications with physical insight. We believe it provides a solution to some aspects of the trans-Planckian issue in de Sitter  since it allows, in particular, for a numerical implementation of the basic equations of QFT on a grid in physical momentum with no need for adding new degrees of freedom as expansion proceeds. Furthermore, it opens the possibility of performing numerical calculations without being limited by the number of e-folds and thus offers a possible way to study various nonperturbative issues in de Sitter space.
Let us finally mention that the $p$-representation is also of relevance in the context of (analog) black-hole physics and Hawking radiation \cite{Brout:1995rd}, where it naturally arises in discussing dispersive and dissipative effects in Lorentz violating theories \cite{Busch:2012ne,ABP}. 

We discuss in detail the physical momentum representation of the original QFT equations in the {\it in-in}, or closed-time-path formalism in Sec. \ref{sec:prep}. We show, in Sec. \ref{sec:diag}, how standard diagrammatic rules for the calculation of two-point vertex functions can be systematically formulated in the $p$-representation. We then illustrate its usefulness when adopting various approximation schemes, such as the perturbative loop expansion and the nonperturbative $1/N$-expansion, discussed in Secs. \ref{sec:loop} and \ref{sec:N} respectively. Finally, we point out in Sec. \ref{sec:2PI} that the approach is particularly suited for nonperturbative (resummed) approximation schemes based on the 2PI formalism. Additional material concerning the reformulation in the $p$-representation of the {\it in-in} closed contour integral, of higher order correlation and vertex functions, 
and of the auxiliary field formulation of the $1/N$ expansion are presented in the Appendices.

\section{$p$-representation of de Sitter correlators}
\label{sec:prep}

We consider a scalar field $\varphi(x)$ in the expanding Poincar\'e patch of de Sitter space with $D=d+1$ dimensions and with Hubble scale $H=1$. The line element is given by
\bea
 ds^2&=&a^2(\eta)\left(-d\eta^2+d\bX^2\right)\nn
 &=&-(1-\bx^2)dt^2-2\bx\cdot d\bx \,dt+d\bx^2\,.
\eea
In the first line, we used the conformal time $\eta$ and comoving spatial coordinates $\bX$, referred to as comoving coordinates in the following. In the second line, we used the cosmological time $t$ and the Lema\^{\i}tre-Painlev\'e-Gullstrand (or physical) spatial coordinates $\bx$, hereafter named PG coordinates. These coordinate systems are related through $\bx=a(\eta)\bX$ and $a(\eta)=-1/\eta=e^t$ with $t\in \mathbb{R}$. 

The comoving coordinates exhibit the homogeneous and expanding character of de Sitter space, whereas the PG coordinates establish that it is also stationary but inhomogeneous. These two facets of de Sitter space are at the very origin of the $p$-representation. References \cite{Busch:2012ne,ABP} present a detailed discussion of the group theoretical foundation of the latter, which is related to the affine subgroup of the de Sitter group\footnote{We warn the interested reader about the different notations used here and in \cite{Busch:2012ne,ABP}: here, we use lower (upper) case letters for physical (comoving) variables, which is the opposite convention of that used in~\cite{Busch:2012ne,ABP}.}.

\subsection{Two-point correlators}

Just like general nonequilibrium quantum systems, quantum field theories on FRW geometries are most conveniently formulated in the so-called Schwinger-Keldysh---also dubbed {\it in-in}---formalism \cite{Schwinger:1960qe,Bakshi:1962dv,Keldysh:1964ud,Chou:1984es,Calzetta:1986cq,Berges:2004vw}, that is on a closed contour $\C$ in the time coordinate; see e.g. \cite{Calzetta:1986ey,Ramsey:1997qc,Tranberg:2008ae}. The appropriate contour for conformal time is depicted in Fig. \ref{fig:etapath} and is discussed in Appendix \ref{appsec:contours}. The various \begin{figure}[h!]  
\epsfig{file=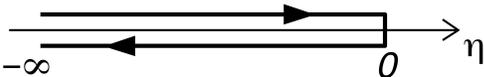,width=6.5cm}
 \caption{\label{fig:etapath} 
The closed path $\C=\C^+\cup\C^-$ in conformal time $x^0=\eta$. The forward (upper) branch $\C^+$ goes from $-\infty$ to $0^-$ and the backward (lower) branch $\C^-$ goes back from $0^-$ to $-\infty$.}
\end{figure}
components of $n$-point correlators are described by means of time-ordered products of field operators on the contour. For instance the two-point function $G(x,x')=\langle T_\C\varphi(x)\varphi(x')\rangle$, where $T_\C$ denotes time ordering along the contour $\C$, encodes both the statistical and spectral correlators\footnote{Here, $\{A,B\}=AB+BA$ and $[A,B]=AB-BA$.} $F(x,x')={1\over2}\langle\{\varphi(x),\varphi(x')\}\rangle$ and $\rho(x,x')=i\langle[\varphi(x),\varphi(x')]\rangle$:
\beq
 G(x,x')=F(x,x')-\frac{i}{2}{\rm sign}_\C(x^0-x^{\prime0})\rho(x,x')\,,
\eeq
where the sign function is to be understood on the contour $\C$; see Appendix \ref{appsec:contours}.
In the rest of this subsection, we consider the statistical function $F$. Everything we write equally applies to the spectral function $\rho$

De Sitter invariance ensures that $F(x,x')$ only depend on the invariant distance $z(x,x')$. In the comoving coordinate system $x=(\eta,\bX)$, 
\beq
 z(x,x')=\frac{\eta^2+\eta^{\prime2}-(\bX-\bX')^2}{2\eta\eta'}.
\eeq
When using these coordinates, it proves convenient to introduce conformally rescaled quantities, such as the field $\phi(x)=a^{d-1\over2}(\eta)\varphi(x)$ and its correlators. One has for instance 
\beq
\label{eq:rep1}
 F(x,x')=\left[a(\eta)a(\eta')\right]^{-{d-1\over2}}F_c(\eta,\eta',|\bX-\bX'|)
\eeq
where $F_c(\eta,\eta',|\bX-\bX'|)={1\over2}\langle \{\phi(x),\phi(x')\}\rangle$. Introducing spatial comoving momentum variables, one writes, with the notation $\int_{\bK}\equiv\int\frac{d^dK}{(2\pi)^d}$,
\beq
 F_c(\eta,\eta',|\bX-\bX'|)=\int_{\bK,\bK'} e^{i\bK\cdot\bX+i\bK'\cdot\bX'}\bar F_c(\eta,\eta',\bK,\bK').
\eeq
Exploiting spatial homogeneity, one gets
\beq
\label{eq:comcons}
 \bar F_c(\eta,\eta',\bK,\bK')=(2\pi)^d\delta^{(d)}(\bK+\bK')\tilde F_c(\eta,\eta',K),
\eeq
where
\beq
 \tilde F_c(\eta,\eta',K)=\int d^d S\, e^{-i\bK\cdot\bS}F_c(\eta,\eta',|\bS|).
\eeq
\Eqn{eq:comcons} simply expresses the conservation of comoving momentum and is valid in any FRW geometry with flat spatial sections.

Let us now see how this relates to PG coordinates. The invariant distance reads
\beq
 z(x,x')=\cosh(\Delta t)-{1\over2}\left(e^{-{1\over2}\Delta t}\bx -e^{{1\over2}\Delta t}\bx'\right)^2,
\eeq
where $\Delta t=t-t'$. We conclude that the two-point correlator can be written as a function of two variables
\bea
\label{eq:rep2}
 F(x,x')&=&F_P(\Delta t,|e^{-{1\over2}\Delta t}\bx -e^{{1\over2}\Delta t}\bx'|)\nn
 &=&\int_{\bp,\bp'}e^{i\bp\cdot\bx+i\bp'\cdot\bx'}\bar F_P(\Delta t,\bp,\bp'),
\eea
where we introduced physical momentum variables in the second line. Exploiting the fact that the dependence of the two-point function on the spatial coordinates is only through the combination $|e^{-{1\over2}\Delta t}\bx -e^{{1\over2}\Delta t}\bx'|$, one easily concludes that the Fourier transform reads
\beq
\label{eq:rigid}
 \bar F_P(\Delta t,\bp,\bp')=(2\pi)^d\delta^{(d)}\left(e^{{1\over2}\Delta t}\bp+e^{-{1\over2}\Delta t}\bp'\right)\tilde F_P(p,p')
\eeq
with
\beq
\label{eq:rigid2}
 \tilde F_P(p,p')=\int d^d s e^{-i\,e^{{1\over2}\Delta t}\bp\cdot\bs}F_P(\Delta t,|\bs|)\,.
\eeq
\Eqn{eq:rigid}  expresses the conservation and redshift of physical momentum, which is nothing but the conservation of comoving momentum $\bK=\bp e^t=-\bp'e^{t'}=-\bK'$. Here, we have used the fact that the integral in \eqn{eq:rigid2} is clearly a function of $\Delta t$ and $p=|\bp|$, which we can trade for $p$ and $p'=pe^{\Delta t}$.

Combining the two equivalent representations \eqn{eq:rep1} and \eqn{eq:rep2} of $F$, we conclude that 
\beq
 \bar F_P(\Delta t,\bp,\bp')=\left[a(\eta)a(\eta')\right]^{{d+1}\over2}\bar F_c(\eta,\eta',a(\eta)\bp,a(\eta')\bp'),
\eeq
and 
\beq
 \tilde F_P(p,p')=\left[a(\eta)a(\eta')\right]^{1/2}\tilde F_c(\eta,\eta',a(\eta)p).
\eeq
Introducing
\beq
\label{eq:tildehat}
 \hat F(p,p')=\sqrt{pp'}\tilde F_P(p,p'),
\eeq
we get the $p$-representation of the two-point function \cite{Busch:2012ne,ABP}
\beq
\label{eq:prep}
 \tilde F_c(\eta,\eta',K)=\frac{1}{K}\hat F(p,p')
\eeq
with $p=-K\eta$ and $p'-K\eta'$, the physical momenta at times $\eta$ and $\eta'$ respectively. This relation expresses the fact that the comoving representation of de Sitter correlators has the scaling property $\tilde F_c(\eta,\eta',K)=\tilde F_c(K\eta,K\eta',1)/K$. As mentioned previously \Eqn{eq:prep} applies to the spectral function $\rho$ as well. For what concerns the calculation of two-point correlators, this effectively reduces the number of independent variables from three to two.

Finally, we notice that in going from the comoving representation to the $p$-representation, the closed contour $\C$ in conformal time is turned into a closed contour $\hat\C$ in physical momentum, as depicted on Fig. \ref{fig:ppath} and discussed in Appendix \begin{figure}[h!]  
\epsfig{file=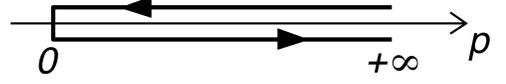,width=6.5cm}
 \caption{\label{fig:ppath} 
The closed path $\hat\C=\hat\C^+\cup\hat\C^-$ in the momentum variable $p=-K\eta$. The upper branch $\hat\C^+$ goes from $+\infty$ to $0^+$ and the lower branch $\hat\C^-$ goes back from $0^+$ to $+\infty$.}
\end{figure}
\ref{appsec:contours}. In fact one can grab the statistical and spectral component of the two-point function in the following propagator on the momentum contour:
\beq
\label{eq:repcontourp}
 \hat G(p,p')=\hat F(p,p')-\frac{i}{2}{\rm sign}_{\hat\C}(p-p')\hat\rho(p,p')
\eeq
where, as before, the sign function is to be understood on the contour; see Appendix \ref{appsec:contours}. We have, in particular, ${\rm sign}_\C(\eta-\eta')={\rm sign}_{\hat\C}(p-p')$.

\subsection{Schwinger-Dyson equations}

We now exploit the above considerations to rewrite the SD equations for the two-point functions in the $p$-representation. We first define the covariant inverse propagator on the closed time contour $G^{-1}$ as
\beq 
\label{eq:inverse}
 \int_z G^{-1}(x,z)G(z,x')=\delta^{(D)}(x,x')
\eeq
with $\int_z\equiv\int d^Dz\sqrt{-g(z)}=\int_\C dz^0\int d^dz\sqrt{-g(z)}$, where the time integral runs along the contour $\C$ and with
\beq
 \delta^{(D)}(x,y)=\frac{\delta^{(D)}(x-y)}{\sqrt{-g(x)}}=\frac{\delta_\C(x^0-y^{0})\delta^{(d)}(\bx-\by)}{\sqrt{-g(x)}}
\eeq
the covariant Dirac distribution on the contour, defined as $\int_z\delta^{(D)}(x,z)f(z)=f(x)$, see Appendix \ref{appsec:contours}. 

Schwinger-Dyson equations are obtained by introducing the covariant self-energy as
\beq
 G^{-1}(x,x')=G_0^{-1}(x,x')-\Sigma(x,x')
\eeq
where the covariant free inverse propagator is given by the quadratic part of the classical action $S[\varphi]$:
\beq
\label{eq:freeprop}
 iG_0^{-1}(x,x')=\left.\frac{\delta_c^2S[\varphi]}{\delta\varphi(x)\delta\varphi(x')}\right|_{\varphi=0}=(\square_x-m_{\rm dS}^2)\delta^{(D)}(x,x'),
\eeq
where 
\beq
\label{eq:covfuncder}
 {\delta_c\over\delta\varphi(x)}\equiv {1\over \sqrt{-g(x)}}{\delta\over\delta\varphi(x)}
\eeq
defines a covariant functional derivative. To fix the ideas we choose here, in the second equality, the standard form of the inverse propagator of a scalar field with standard kinetic and mass terms. Here,
\beq
\label{eq:laplace}
 \square_x\equiv\frac{1}{\sqrt{-g(x)}}\partial_\mu\sqrt{-g(x)}g^{\mu\nu}\partial_\nu
\eeq
is the covariant Laplace operator and 
\beq
\label{eq:lamasse}
 m_{\rm dS}^2=m^2+\xi R=m^2+d(d+1)\xi
\eeq
is the effective square mass with $m$ the tree-level mass and $\xi$ the coupling to curvature $R=d(d+1)$. 

Extracting a possible local contribution to the self-energy\footnote{We extract a local term to take into account possible local (tadpole diagrams) contributions to the self-energy. It is to be emphasized that the local contribution to the self-energy is modified by the renormalization of UV divergences. In particular, in $D=4$, one expects an additional contribution $\sim \square_x\delta^{(4)}(x,x')$ corresponding to field strength renormalization \cite{Brunier:2004sb}.} (note that de Sitter symmetry imposes that the local term $\sigma$ be constant),
\beq
\label{eq:local}
 \Sigma(x,x')=-i\sigma\delta^{(D)}(x,x')+\Sigma_{\rm nl}(x,x'),
\eeq
the SD equation reads
\beq
\label{eq:SD1}
\left[\square_x-M^2\right] \!G(x,x') \!=\!i\delta^{(D)}(x,x')+i\!\int_z\Sigma_{\rm nl}(x,z) G(z,x'),
\eeq
where we defined
\beq
 M^2=m_{\rm dS}^2+\sigma\,.
\eeq

\subsubsection{Comoving representation}

Let us now specify to the comoving representation. As before, we define the inverse comoving propagator with conformal rescaling factors:
\beq
 G^{-1}(x,x')=\left[a(\eta)a(\eta')\right]^{-{d+3\over2}}G_c^{-1}(\eta,\eta',|\bX-\bX'|)
\eeq
such that, in comoving momentum space, \Eqn{eq:inverse} reads
\beq
\label{eq:inversecom}
 \int_\C d\xi\,  \tilde G_c^{-1}(\eta,\xi ,K)\tilde G_c(\xi ,\eta',K)=\delta_\C(\eta-\eta')
\eeq
The conformally rescaled self-energy is defined accordingly:
\beq
\label{eq:sigmacom}
 \Sigma(x,x')=\left[a(\eta)a(\eta')\right]^{-{d+3\over2}}\Sigma_c(\eta,\eta',|\bX-\bX'|).
\eeq
and \Eqn{eq:local} becomes, in comoving momentum space,
\beq
\label{eq:sigmacomov}
 \tilde\Sigma_c(\eta,\eta',K)=-i\sigma a^2(\eta)\delta_\C(\eta-\eta')+\tilde\Sigma_c^{\rm nl}(\eta,\eta',K).
\eeq
where we used $\sqrt{-g(x)}=a^{D}(\eta)$. Finally, \Eqn{eq:SD1} takes the form
\bea
&&\left[\partial_\eta^2+K^2 -\frac{\nu^2-{1\over4}}{\eta^2}\right]\tilde G_c(\eta,\eta',K) =-i\delta_\C(\eta-\eta')\nn
\label{eq:comeq}
&&\hspace{2.5cm}-i\int_\C d \xi \,
\tilde\Sigma_c^{\rm nl}(\eta,\xi ,K)\tilde G_c(\xi ,\eta',K),\nn
\eea
where
\beq
 \nu=\sqrt{\frac{d^2}{4}-M^2}.
\eeq

In order to write the explicit form of the time integrals along the closed contour, we use the standard decomposition of (nonlocal) two-point functions \cite{Berges:2004vw}
\beq
 \tilde G_c(\eta,\eta',K)=\tilde F_c(\eta,\eta',K)-\frac{i}{2}{\rm sign}_\C(\eta-\eta')\tilde \rho_c(\eta,\eta',K)
\eeq
and 
\beq
\label{eq:decselfcom}
 \tilde\Sigma_c^{\rm nl}(\eta,\eta',K)=\tilde\Sigma_c^F(\eta,\eta',K)-\frac{i}{2}{\rm sign}_\C(\eta-\eta')\tilde\Sigma_c^\rho(\eta,\eta',K).
\eeq
It is a straightforward exercise to show that \cite{Berges:2004vw}
\bea
 &&\hspace{-.5cm}i\int_\C d\xi\,  A(\eta,\xi )B(\xi ,\eta')=\nn
 &&\int_{-\infty}^\eta\!\! d \xi \,A_\rho(\eta,\xi )B_F(\xi ,\eta')-\int_{-\infty}^{\eta'}\!\! d \xi \,A_F(\eta,\xi )B_\rho(\xi ,\eta')\nn
\label{eq:exo}
 &&-{i\over2}{\rm sign}_\C(\eta-\eta')\int_{\eta'}^\eta d \xi \,A_\rho(\eta,\xi )B_\rho(\xi ,\eta'),
\eea
so that the SD equations on the time contour read \cite{Tranberg:2008ae}
\bea
&&\left[\partial_\eta^2+K^2 -\frac{\nu^2-{1\over4}}{\eta^2}\right] \tilde F_c(\eta,\eta',K) \nn
&&\hspace{1.9cm}= \int_{-\infty}^{\eta'}\!\! d \xi \,  \tilde \Sigma_c^{F}(\eta,\xi ,K) \tilde \rho_c(\xi ,\eta',K) \nonumber \\
&&\hspace{1.9cm}- \int_{-\infty}^{\eta}\!\! d \xi \,\tilde \Sigma_c^{\rho}(\eta,\xi ,K) \tilde F_c(\xi ,\eta',K),
\label{eq:F1}
\eea
\bea
&&\left[\partial_\eta^2+K^2  -\frac{\nu^2-{1\over4}}{\eta^2}\right] \tilde\rho_c(\eta,\eta',K) \nn
&&\hspace{1.6cm}=-\int_{\eta'}^{\eta}  d \xi \,
\tilde\Sigma_c^{\rho}(\eta,\xi ,K)\tilde\rho_c(\xi ,\eta',K).
\label{eq:rho1}
\eea
These are nonlinear integro-differential equations. They are nonlocal and causal since they involve memory integrals over the whole past history of the system. 

Being second order in time, these equations must be supplemented by initial data for the functions $\tilde F_c$ and $\tilde\rho_c$ and their first two derivatives e.g. at $\eta=\eta'\to-\infty$. The initial conditions for the spectral function are given by equal-time commutation relations: $\tilde\rho_c(\eta,\eta,K)=\partial_\eta\partial_{\eta'}\tilde\rho_c(\eta,\eta',K)|_{\eta=\eta'}=0$ and $\partial_\eta\tilde\rho_c(\eta,\eta',K)|_{\eta=\eta'}=1$. The statistical function
contains the information about the (quantum) state of the system. Renormalizability (or Hadamard conditions) select the so-called Bunch-Davies vacuum state as the only viable de Sitter invariant state. In the infinite past (subhorizon limit), the latter reduces to the corresponding Minkowsky vacuum state of the interacting theory, which can still be a quite complicated state. However, if we define the interacting theory by means of an adiabatic switch on of the interaction from the infinite remote past, we may identify the state at $\eta\to-\infty$ as the free Bunch-Davies vacuum, characterized by 
\bea
 \left.\tilde F(\eta,\eta',K)\right|_{\eta=\eta'\to-\infty}&=&{1\over2K},\nn
 \left.\partial_\eta\tilde F(\eta,\eta',K)\right|_{\eta=\eta'\to-\infty}&=&0,\\
 \left.\partial_\eta\partial_{\eta'}\tilde F(\eta,\eta',K)\right|_{\eta=\eta'\to-\infty}&=&{K\over2}.\nonumber
\eea

As discussed in the introduction, the SD equations in the comoving representation, Eqs. \eqn{eq:F1} \eqn{eq:rho1}, have a very similar structure as nonequilibrium evolution equations for a scalar field in flat space-time: the only place where expansion enters is the time-dependent mass term $\propto 1/\eta^2$. In this representation, the calculation of de Sitter correlators is fully expressed as an initial value problem. Exploiting the spatial homogenetity and isotropy of de Sitter geometry in comoving coordinates, the problem effectively reduces to the calculation of a ($1+1$)-dimensional nonequilibrium two-point correlator. 

We can readily see the difficulties mentioned in the introduction on Eqs. \eqn{eq:F1}-\eqn{eq:rho1}. If one chooses a discretization on a (fixed) grid in comoving coordinates, the growing mass term eventually becomes larger than the comoving momentum cutoff. It thus requires an initially extremely fine lattice in order to resolve the inverse mass during a large number of e-folds. At the same time one wants an as large as possible spatial volume in order to correctly describe IR physics. In practice, this approach is bounded to only a few e-folds \cite{Tranberg:2008ae}. 

Moreover, as already emphasized, such a discretization does not have a continuum limit in $D=4$. A more appropriate choice in this respect is to discretize the system in proper physical coordinates. However, in that case, the number of comoving modes $K$ involved in the simulation increases with time. The will to correctly describe IR physics (which requires a large volume) for a long time (which eventually requires a large number of degrees of freedom) leads to a similar difficulty as in the previous case. Another issue in that case is that one needs to supplement the evolution equations with an {\it ad hoc} specification of how to initialize the new degrees of freedom which constantly enter the system.

Let us now discuss how SD equations can be formulated in the $p$-representation and show how this solves the above issues.

\subsubsection{$p$-representation}

As already mentioned the contour $\C$ in conformal time can be traded for a contour $\hat\C$ in momentum, as depicted in Fig. \ref{fig:ppath} above. We first define the $p$-representation of the inverse propagator as
\beq
 \tilde G_c^{-1}(\eta,\eta',K)=K^3\hat G^{-1}(p,p'),
\eeq
with $p=-K\eta$ and $p'=-K\eta'$, such that \Eqn{eq:inversecom} becomes
\beq
 \int_{\hat\C} ds\,\hat G^{-1}(p,s)G(s,p')=\delta_{\hat\C}(p-p'),
\eeq
where $\delta_\C(\eta)$ is the delta function on $\hat\C$, defined such that $\int_{\hat\C}ds\,\delta_{\hat\C}(p-s)f(s)=f(p)$, see Appendix \ref{appsec:contours}. Notice, in particular, the relation $\delta_\C(\eta-\eta')=-K\delta_{\hat\C}(p-p')$.

Assuming that the self-energy scales as the inverse propagator\footnote{There is a freedom in the choice of the $p$-representation of the inverse propagator. For instance, in Ref. \cite{ABP}, the covariant inverse propagator and propagator are treated on an equal footing. In the present notations, the authors of \cite{ABP} thus introduce the following function, see \Eqn{eq:rep1}: $\Sigma(x,x')=[a(\eta)a(\eta')]^{-{d-1\over2}}\Sigma_{\rm ABP}(\eta,\eta',|\bX-\bX'|)$, which admits the $p$-representation, see \Eqn{eq:prep} $\tilde\Sigma_{\rm ABP}(\eta,\eta',K)=\hat\Sigma_{\rm ABP}(p,p')/K$. With this choice, the convolutions in the physical momentum variable, e.g. in Eqs. \eqn{eq:SDp1}-\eqn{eq:SDp2}, involve a nontrivial measure $\int ds/s^2$. Here, we treat the covariant inverse propagators and propagators differently. Our choice is such that convolutions in the physical momenta involve a trivial measure. Our self-energy is related to that of  \cite{ABP} by $\hat\Sigma_{\rm ABP}(p,p')=(pp')^2\hat\Sigma(p,p')$.},
\beq
\label{eq:selfscale}
 \tilde\Sigma_c(\eta,\eta',K)=K^3\hat \Sigma(p,p'),
\eeq
one can write 
\beq
\label{eq:sigmalocprep}
 \hat\Sigma(p,p')=i\sigma{\delta_{\hat\C}(p-p')\over p^2}+\hat\Sigma_{\rm nl}(p,p'),
\eeq
where the nonlocal contribution is defined as $\tilde\Sigma_c^{\rm nl}(\eta,\eta',K)=K^3\hat \Sigma_{\rm nl}(p,p')$, and the SD equation can be rewritten fully in the $p$-representation as
\bea
\left[\partial_p^2+1 -\frac{\nu^2-{1\over4}}{p^2}\right]\hat G(p,p') &=&i\delta_{\hat\C}(p-p')\nn
&&\hspace{-1.3cm}+\,i\!\int_{\hat\C}\!  d s \, \hat\Sigma_{\rm nl}(p,s )\hat G(s ,p'),
\eea
As before, the contour integral can be written explicitly. We write 
\bea
 \hat G(p,p')\!\!&=&\!\!\hat F(p,p')-\frac{i}{2}{\rm sign}_{\hat\C}(p-p')\hat\rho(p,p'),\\
 \label{eq:decselfp}
 \hat \Sigma_{\rm nl}(p,p')\!\!&=&\!\!\hat\Sigma_F(p,p')-\frac{i}{2}{\rm sign}_{\hat\C}(p-p')\hat\Sigma_\rho(p,p'),
\eea
and similarly for any nonlocal two-point function on the contour $\hat\C$. \Eqn{eq:exo} becomes
\bea
 &&\hspace{-.5cm}-i\int_{\hat\C} ds\,  A(p,s )B(s ,p')=\nn
 &&\int_{p}^\infty\!\! d s \,A_\rho(p,s )B_F(s ,p')-\int_{p'}^{\infty}\!\! d s \,A_F(p,s )B_\rho(s ,p')\nn
\label{eq:exo2}
 &&-{i\over2}{\rm sign}_{\hat\C}(p-p')\int^{p'}_p d s \,A_\rho(p,s )B_\rho(s ,p'),
\eea
and the SD equations thus read
\bea
\label{eq:SDp1}
\left[\partial_p^2+1-\frac{\nu^2-{1\over4}}{p^2}\right] \hat F(p,p')&=&  \int^{\infty}_{p'}\!\!\! d s \,  \hat\Sigma_{F}(p,s ) \hat\rho(s ,p')\nonumber \\
&-& \int^{\infty}_{p}\!\!\! d s \,\hat\Sigma_{\rho}(p,s ) \hat F(s ,p'),\nn
\\
\label{eq:SDp2}
\left[\partial_p^2+1-\frac{\nu^2-{1\over4}}{p^2}\right] \hat\rho(p,p')
&=&-\int^{p'}_{p}\!  d s \,\hat\Sigma_{\rho}(p,s )\hat\rho(s ,p').\nn
\eea
The ``initial'' data are to be specified at $p=p'\to\infty$. Commutation relations imply $\hat\rho(p,p')|_{p=p}=\partial_p\partial_{p'}\hat\rho(p,p')|_{p=p'}=0$ and $\partial_p\hat\rho(p,p')|_{p=p'}=-1$ for the spectral function and the choice of the free Bunch-Davies vacuum at large momentum---keeping in mind an adiabatic switching on of the interaction---means
\bea
 \left.\hat F(p,p')\right|_{p=p'\to\infty}&=&{1\over2}\nn
 \left.\partial_p\hat F(p,p')\right|_{p=p'\to\infty}&=&0\\
 \left.\partial_p\partial_{p'}\hat F(p,p')\right|_{p=p'\to\infty}&=&{1\over2}\nonumber
\eea
for the statistical correlator.

Eqs. \eqn{eq:SDp1} \eqn{eq:SDp2} generalize the evolution equations of a free field in the $p$-representation introduced in \cite{Busch:2012ne}.  They provide an important simplification as compared to the comoving formulation in that the problem is reduced to an effective ($0+1$)-dimensional, quantum-mechanical-like problem, with two-point correlators depending only on two momentum/time variables, instead of an effective ($1+1$)-dimensional problem with two time an one momentum variables in the comoving representation. This is of particular importance for numerical investigations of SD equations on de Sitter space using nonequilibrium techniques. 
  
It is remarkable that in the $p$-representation the time evolution is replaced by a momentum evolution and, as a consequence, for given self-energy kernels $\hat\Sigma_{F,\rho}$, the SD equations turn into integro-differential flow equations in physical momentum with a second order derivative. This is for instance the case in ordinary perturbation theory, where the self-energies at a given order are given functions of the free propagators. In such a case, the calculation of the propagator for a given momentum only involves higher momenta and Eqs. \eqn{eq:SDp1} \eqn{eq:SDp2} describe how the integration of higher momenta builds up lower momenta correlators. However, this is no longer the case for self-consistent approximation schemes such as those based on the 2PI formalism, where self-energies involve momentum (loop) integrals of the full propagators themselves and thus depend on the latter at all momenta. In such a case one can envisage obtaining a nonperturbative solution by means of iterative techniques \cite{Gautier}.

\section{Diagrammatic rules}
\label{sec:diag}

The considerations of the previous section rely on the scaling assumption \eqn{eq:selfscale}, which ensures that the $p$-representation closes. However the self-energy is generated by the field (self-)interactions and one has to check that this scaling relation is actually satisfied whenever the scaling \eqn{eq:prep} for the correlator is true. To do so, in the next subsection, we shall analyze the diagrammatic representation of the self-energy and derive Feynman rules in the $p$-representation. The case of higher order correlation or vertex functions is briefly discussed in Appendix \ref{appsec:higher}. Finally we give some explicit examples of self-energies in the $p$-representation in both perturbative and nonperturbative approximation schemes.

\subsection{Two-point functions}

We start from the usual diagrammatic rules for the self-energy $\Sigma(x,x')$ in the covariant formulation. We are thus concerned with connected one-particle-irreducible (1PI) diagrams with two external (amputated i.e. not associated to a propagator) legs. We do not need to specify any particular interaction term. We assume nonderivative---but otherwise arbitrary---polynomial interactions. Consider a given 1PI diagram: each internal line contributes a $G(z_i,z_i')$ with $z_i$ and $z_i'$ the endpoints of the $i$th line; each vertex with $n$ legs contributes a $\prod_{k=1}^{n-1}\delta^{(D)}(z^j_k,z^j_{k+1})$ with $z^j_1,\ldots,z^j_n$ the space-time coordinates associated with each leg of the $j$th vertex; finally all coordinates associated with vertices must be integrated over with the covariant measure, $\int_z$, except for the two coordinates $x$ and $x'$ associated with the two external legs. For a diagram with $I$ internal lines there are $2I$ space-time coordinates to be integrated over. 

Let us now move on to the comoving representation in momentum space, i.e. to the diagrammatic rules for $\tilde\Sigma_c(\eta,\eta',K)$. First, there is now an overall factor $[a(\eta)a(\eta')]^{d+3\over2}=(\eta\eta')^{-{d+3\over2}}$ from the definition of the conformally rescaled self-energy, \Eqn{eq:sigmacom}. To each internal line of the diagram under consideration is associated a comoving momentum $\bQ_i$, to be integrated over, and two conformal time endpoints $\eta_i$ and $\eta_i'$. Each such line contributes a factor $(\eta_i\eta_i')^{d-1\over2}\tilde G_c(\eta_i,\eta_i',Q_i)$. Each vertex with $n$ legs contributes a comoving momentum conservation factor $(2\pi)^d\delta^{(d)}\left(\sum_{i=1}^n \bQ_i\right)$, where the sum runs over all momenta entering the vertex. One of these Dirac factors ensures the total comoving momentum conservation, as e.g. in \Eqn{eq:comcons}, and is extracted in the definition of $\tilde\Sigma_c$. Therefore, for a diagram with $V$ vertices, there are $V-1$ momentum conservation factors. The number of independent momentum integrations (loops) is thus $L=I-V+1$, the usual relation. Each vertex not attached to an external endpoint contributes an integral over a conformal time variable $\int_\C d\eta_i a^D(\eta_i)$. There are $V-2$ such vertices for nonlocal contributions to the self-energy (local terms are to be treated separately).

We next translate the above rules to the $p$-representation. To do so, we rescale all---internal and external---conformal time variables as well as internal momenta with the external comoving momentum $K$: $p=-K\eta$, $p'=-K\eta'$ for time variables associated with external legs, $p_i=-K\eta_i$, $p_i'=-K\eta_i'$ for times variables associated with internal vertices, and $\bQ_i=K\bq_i$ for internal momenta. Each line factor then reads
\beq
\label{eq:linefactor}
 d^dQ_i\,(\eta_i\eta_i')^{d-1\over2}\tilde G_c(\eta_i,\eta_i',Q_i)=d^dq_i\,(p_ip_i')^{d-1\over2}\frac{\hat G\!\left(q_ip_i,q_ip_i'\right)}{q_i}
\eeq
We see that although some endpoints $\eta_i,\eta_i'$ of some lines are actually equal to the external ones $\eta,\eta'$, the line factors can be entirely written in terms of $p,p'$ thanks to the fact that only ratios, e.g. $\eta/\eta_i=p/p_i$, occur. Now, each of the $V-1$ comoving momentum conservation terms
\beq
 \delta^{(d)}\left(\sum_i \bQ_i\right)=K^{-d}\delta^{(d)}\left(\sum_i \bq_i\right)
\eeq
contributes a factor $K^{-d}$ and each integral over conformal times  associated to internal vertices (not attached to an external leg)
\beq
 \int_\C d\eta_i a^D(\eta_i)=\int_\C \frac{d\eta_i}{(-\eta_i)^D}=-K^d\int_{\hat\C} \frac{dp_i}{p_i^D}
\eeq
contributes a factor\footnote{Here, the minus sign on the right-hand side is to be included in the diagrammatic rule for integrating over internal vertices. It simply reflects the orientation of the momentum contour.} $K^d$. 

A nonlocal contribution to $\Sigma$ is such that the two external legs are not attached to the same vertex. A diagram with $V$ vertices has thus $V-2$ internal vertices and contributes a term 
\beq
\label{eq:factorover}
 \frac{K^{d(V-2)}}{K^{d(V-1)}}\times(\eta\eta')^{-{d+3\over2}}=K^3\times (pp')^{-{d+3\over2}}
\eeq
times a function of $p$ and $p'$ only. We emphasize that this is independent of the number of vertices $V$ and thus of the particular diagram under consideration. This demonstrates the $K^3$ scaling of self-energy, \Eqn{eq:selfscale}, at any order of perturbation theory as a consequence of the scaling \eqn{eq:prep}. 

For a local contribution to the self-energy, both external lines are attached to the same vertex and there are thus $V-1$ internal vertices and an extra $\delta_\C(\eta-\eta')/a^D(\eta)$. Altogether one gets an overall factor
\beq
 \frac{K^{d(V-1)}}{K^{d(V-1)}}\times\frac{(-\eta)^{D}\delta_\C(\eta-\eta')}{(\eta\eta')^{{d+3\over2}}}=-K^3\times \frac{\delta_{\hat\C}(p-p')}{p^2}
\eeq
as required for the $p$-representation, see \eqn{eq:sigmacomov} and \eqn{eq:sigmalocprep}.

Finally, we emphasize that the previous analysis shows that the diagrammatic rules in the $p$-representation are the same as those in the comoving representation with the generic replacements $\bQ\to\bq$ for all momenta (including the external one for which one has $\bK\to\be$), $-\eta\to p$ for all time variables and $\tilde G_c\to\hat G/q$ for all propagator lines, see \Eqn{eq:linefactor}. In the next subsections, we give explicit examples for various approximation schemes. The diagrammatic rules for higher correlation and vertex functions in the $p$-representation are discussed in Appendix \ref{appsec:higher}.

\section{Loop expansion}
\label{sec:loop}

We illustrate the above considerations for an $O(N)$ theory with quartic coupling. With the definitions \eqn{eq:laplace} and \eqn{eq:lamasse}, the classical action reads
\beq
\label{eq:classical}
 {\cal S}[\varphi]=\int_x\left\{{1\over2}\varphi_a\left(\square-m_{\rm dS}^2\right)\varphi_a-\frac{\lambda}{4!N}(\varphi_a\varphi_a)^2\right\},
\eeq
where $a=1,\ldots,N$ and a summation over repeated indices is understood. We consider the symmetric phase, $\langle\varphi_a\rangle=0$, for which the propagator and self-energy are diagonal: $G_{ab}=\delta_{ab}G$ and $\Sigma_{ab}=\delta_{ab}\Sigma$.

As a first example of an approximation scheme, we consider the standard loop expansion which, in the case under consideration, is equivalent to a coupling expansion. We write the formal series
\beq
 \Sigma=\Sigma^{(1)}+\Sigma^{(2)}+\ldots
\eeq
where $\Sigma^{(n)}\sim{\cal O}(\lambda^n)$ is the $n$-loop order contribution.

\subsection{One loop}

\begin{figure}[h!]  
\epsfig{file=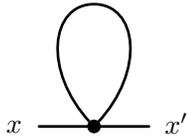,width=2.5cm}
 \caption{\label{fig:oneloop} 
The one-loop tadpole contribution to the self-energy $\Sigma(x,x')$. The black dot denotes an interaction vertex and the line in the loop represents the free propagator $G_0(x,x)$. A similar diagram describes local contributions in the $1/N$ expansion. In that case, the line represents the leading order (large-$N$) propagator $G(x,x)$.}
\end{figure}

Let us first recall the result in the covariant formulation. At one-loop order there is only a local contribution given by the tadpole diagram of Fig. \ref{fig:oneloop}
\beq
 \Sigma^{(1)}(x,x')=-i\sigma^{(1)}\delta^{(D)}(x,x'),
\eeq
with
\beq
\label{eq:s1}
 \sigma^{(1)}=gG_0(x,x).
\eeq
where we defined $g=\lambda(N+2)/6N$ and where $G_0$ denotes the free (covariant) propagator; see \Eqn{eq:freeprop}. The de Sitter symmetry group guarantees that $G_0(x,x)$ only depends on the invariant distance $z(x,x)=1$. The mass shift $\sigma^{(1)}$ is thus a constant. Applying the diagrammatic rules directly in the $p$-representation, we obtain
\beq
\label{eq:s11}
 \sigma^{(1)}=g\int_\bq p^{d-1}\frac{\hat G_0(qp,qp)}{q}=g\int_\bq\frac{\hat F_0(q,q)}{q},
\eeq
where we made the change of variable $\bq\to\bq/p$ and used the representation \eqn{eq:repcontourp} for the free propagator on the contour in the second equality. This is readily seen to coincide with the above expression\footnote{Note that the divergent integral in \Eqn{eq:s11} needs to be regulated. Note also that a cutoff on comoving momenta would lead to a time dependent result and would thus be inconsistent with the de Sitter symmetry.} \Eqn{eq:s1}, using Eqs. \eqn{eq:rep2}, \eqn{eq:rigid} and \eqn{eq:tildehat}. 

\subsection{Two loop}

\begin{figure}[h!]  
\epsfig{file=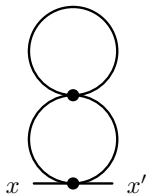,width=2cm}
 \caption{\label{fig:twolooplocal} 
The local two-loop contribution to $\Sigma(x,x')$.}
\end{figure}

\begin{figure}[h!]  
\epsfig{file=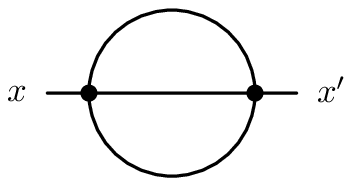,width=4cm}
 \caption{\label{fig:twoloop} 
The nonlocal two-loop contribution to $\Sigma(x,x')$.}
\end{figure}

At two loop there is both a local and a nonlocal contribution. They are depicted in Figs. \ref{fig:twolooplocal} and \ref{fig:twoloop} respectively:
\beq
 \Sigma^{(2)}(x,x')=-i\sigma^{(2)}\delta^{(D)}(x,x')+\Sigma^{(2)}_{\rm nl}(x,x').
\eeq
The local contribution is a first example with an internal vertex. It reads, in the $p$-representation,
\beq
 \sigma^{(2)}=ig^2\int_{\hat\C}\frac{ds}{s^D}\int_\bk(ps)^{d-1}\frac{\hat G_0^2(kp,ks)}{k^2}\int_\bq s^{d-1}\frac{\hat G_0(qs,qs)}{q}.
\eeq
Applying the changes of variables $\bq\to\bq/s$, $\bk\to\bk/p$ and $s\to ps$ and using \eqn{eq:s11}, it can be rewritten as
\beq
 \sigma^{(2)}=ig\sigma^{(1)}\int_{\hat\C}\frac{ds}{s^2}\int_\bk \frac{\hat G_0^2(k,ks)}{k^2},
\eeq
which clearly shows that it is indeed a constant, as required by de Sitter symmetry. Writing the integral on the contour $\hat\C$ explicitly using \Eqn{eq:exo2}, one also checks that $\sigma_0^{(2)}$ is real as expected\footnote{An alternative expression is
$$ \sigma^{(2)}=-2g\sigma^{(1)}\int_\bk \int_k^\infty\frac{ds}{s^2}\frac{\hat F_0(k,s)\hat\rho_0(k,s)}{k}.
$$}: 
\beq
 \sigma^{(2)}=-2g\sigma^{(1)}\int_1^\infty\frac{ds}{s^2}\int_\bk \frac{\hat F_0(k,ks)\hat\rho_0(k,ks)}{k^2}.
\eeq

The nonlocal contribution reads, in the covariant representation,
\beq
\label{eq:previous1}
 \Sigma^{(2)}_{\rm nl}(x,x')=g'G_0^3(x,x'),
\eeq
where $g'=-\lambda^2(N+2)/18N^2$, or, in the comoving representation, 
\bea
 \label{eq:previous2}
 &&\hspace{-0.5cm}\tilde\Sigma^{(2)}_{\rm nl}(\eta,\eta',K)=g'(\eta\eta')^{d-3}\\
 &&\,\,\times\int_{\bQ,\bL}\!\!\!\tilde G_{c,0}\left(\eta,\eta',Q\right)\tilde G_{c,0}\left(\eta,\eta',L\right)\tilde G_{c,0}\left(\eta,\eta',R\right)\nonumber
\eea
where $R=|K\be+\bQ+\bL|$ with $\be$ an arbitrary unit vector. 

A direct application of the diagrammatic rules in the $p$-representation gives
\beq
\label{eq:S2}
 \hat\Sigma^{(2)}_{\rm nl}(p,p')\!=\!g'(pp')^{d-3}\!\!\!\int_{\bq,\bl}\!\!\!\frac{\hat G_0\!\left(qp,qp'\right)\!\hat G_0\!\left(lp,lp'\right)\!\hat G_0\!\left(rp,rp'\right)}{qlr}
\eeq
where $r=|\be+\bq+\bl|$. \Eqn{eq:S2} is easily checked to coincide with the previous expressions \eqn{eq:previous1} or \eqn{eq:previous2} when converted in the appropriate representation. The explicit expressions of the component $\Sigma_F^{(2)}$ and $\Sigma_\rho^{(2)}$ are also easily obtained: the product $\hat G_0\hat G_0\hat G_0$ under the integral gives rise to the combinations (keeping the same momentum arguments) $\hat F_0\hat F_0\hat F_0-{3\over4}\hat F_0\hat\rho_0\hat\rho_0$ for $\Sigma_F^{(2)}$ and $3\hat F_0\hat F_0\hat\rho_0-{1\over4}\hat\rho_0\hat\rho_0\hat\rho_0$ for $\Sigma_\rho^{(2)}$.

\section{$1/N$ expansion}
\label{sec:N}

Let us now consider an example of a nonperturbative approximation scheme, the $1/N$ expansion. The latter is a powerful tool to describe nontrivial IR physics in situations where perturbation theory fails. Exact results for IR de Sitter correlators have been recently obtained in the large-$N$ limit \cite{Serreau:2011fu,4pt}, which reveals interesting phenomena such as radiative symmetry restoration, or the generation of anomalous dimensions. We write the formal series in $1/N$
\beq
 \Sigma=\Sigma^{\rm LO}+\Sigma^{\rm NLO}+\ldots
\eeq
where $\Sigma^{\rm LO}\sim {\cal O}(N^0)$, $\Sigma^{\rm NLO}\sim {\cal O}(1/N)$, etc. The diagrammatics of the $1/N$ expansion  in flat space-time is well known \cite{Coleman:1974jh,Root:1974zr,Cooper:1994hr,Aarts:2002dj,Cooper:2004rs}. The generalization to arbitrary background geometry is straightforward in the covariant formulation. We do not recall the derivations here but merely state the results and show how they can be written in the $p$-representation. We follow the notations of Ref. \cite{Aarts:2002dj}.

\subsection{Leading order}

The leading order (LO) contribution is a simple tadpole diagram; see Fig. \ref{fig:oneloop}. It is local and has a similar structure to the one-loop result discussed above. The essential difference is that the tadpole loop is given self-consistently---hence the nonperturbative nature of the approximation scheme---in terms of the full LO propagator $G$. One has
\beq
 \Sigma^{\rm LO}(x,x')=-i\sigma^{\rm LO}\delta^{(D)}(x,x').
\eeq
with
\beq
 \sigma^{\rm LO}=\frac{\lambda}{6}G(x,x).
\eeq
The LO propagator $G$ is defined by [see \Eqn{eq:SD1}]
\beq
 \left(\square-M^2_{\rm LO}\right)G(x,x')=i\delta^{(D)}(x,x')
\eeq
with
\beq
 M^2_{\rm LO}=m_{\rm dS}^2+\sigma^{\rm LO}.
\eeq
The $p$-representation of the LO approximation reads
\beq
 \sigma^{\rm LO}=\frac{\lambda}{6}\int_\bq\frac{\hat F(q,q)}{q}.
\eeq

The diagrammatic $1/N$ expansion can be expressed fully in terms of the LO propagator $G$, which resums the infinite series of so-called daisy and superdaisy tadpole diagrams \cite{Root:1974zr}. Alternatively, it proves convenient to introduce an auxiliary composite field $\chi\propto\varphi_a\varphi_a$ in order to organize the $1/N$ expansion \cite{Coleman:1974jh,Root:1974zr,Cooper:1994hr,Aarts:2002dj,Cooper:2004rs}. We shall follow the first approach here. The auxiliary field formulation is briefly discussed in Appendix \ref{appeq:NLOaux}.

\subsection{Next-to-leading order}

\begin{figure}[h!]  
\epsfig{file=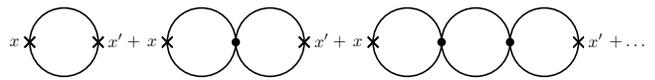,width=8.5cm}
 \caption{\label{fig:bubbles} 
The infinite series of bubble diagrams contributing to the function $I(x,x')$, \Eqn{eq:Ifunc}. The black dots correspond to interaction vertices whereas the crosses denote the endpoints of the function. The elementary bubble is given by the function $\Pi(x,x')$, \Eqn{eq:Pifunc}. Each additional bubble involves a summation of field components and thus comes with a factor $N$, which is compensated by a $1/N$ from the corresponding additional vertex. All such diagrams are thus of the same order in $1/N$.}
\end{figure}

\begin{figure}[h!]  
\epsfig{file=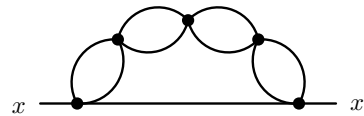,width=5cm}
 \caption{\label{fig:NLO1} 
A typical multiloop diagram contributing to the self-energy $\Sigma(x,x')$ at NLO in the $1/N$ expansion. The latter actually resums all diagrams of similar topology with an arbitrary number of bubbles in the upper part, as described by the function $I(x,x')$; see Fig. \ref{fig:bubbles}.}
\end{figure}

The next-to-leading order (NLO) contribution contains both a local and a nonlocal part:
\beq
\label{eq:NLO1}
 \Sigma^{\rm NLO}(x,x')=-i\sigma^{\rm NLO}\delta^{(D)}(x,x')+\Sigma^{\rm NLO}_{\rm nl}(x,x').
\eeq
The local part is simply given by
\beq
\label{eq:NLOlocal}
 \sigma^{\rm NLO}=\frac{\lambda}{3N}G(x,x)=\frac{2}{N}\sigma^{\rm LO}
\eeq
and the nonlocal part resums the infinite series of diagrams shown in Figs. \ref{fig:bubbles} and \ref{fig:NLO1}. It can be written as
\beq
\label{eq:NLOnl}
 \Sigma^{\rm NLO}_{\rm nl}(x,x')=\frac{\lambda}{3N}G(x,x')I(x,x'),
\eeq
where the function $I$ resums the infinite series of bubble diagrams shown in Fig. \ref{fig:bubbles} through the following integral equation
\beq
\label{eq:Ifunc}
 {I}(x,x')=\Pi(x,x')+i\int_z\Pi(x,z){I}(z,x'),
\eeq
with the elementary one-loop bubble
\beq
\label{eq:Pifunc}
 \Pi(x,x')=-\frac{\lambda}{6}G^2(x,x').
\eeq

The NLO contribution can be expressed in the comoving representation by introducing the conformally rescaled quantities
\beq
 \Pi(x,x')=[a(\eta)a(\eta')]^{-{d+1\over2}}\Pi_c(\eta,\eta',|\bX-\bX'|)
\eeq
and similarly for $I$. One obtains, in comoving momentum space,
\beq
 \tilde\Pi_c(\eta,\eta',K)=-\frac{\lambda}{6}\,(\eta\eta')^{d-3\over2}\!\!\int_\bQ\tilde G_c\left(\eta,\eta',Q\right)\tilde G_c\left(\eta,\eta',R\right),
\eeq
where $R=|K\be+\bQ|$, and 
\beq
 \tilde I_c(\eta,\eta',K)=\tilde\Pi_c(\eta,\eta',K)+i\int_\C d\xi\, \,\tilde\Pi_c(\eta,\xi ,K)\tilde I_c(\xi ,\eta',K).
\eeq
Finally the  nonlocal part of the NLO self-energy reads
\beq
 \tilde\Sigma^{\rm NLO}_{c,{\rm nl}}(\eta,\eta',K)=\frac{\lambda}{3N}\,(\eta\eta')^{d-3\over2}\!\!\int_\bQ\tilde G_c\left(\eta,\eta',Q\right)\tilde I_c\left(\eta,\eta',R\right)\!.
\eeq

Using the methods described in previous sections, it is easy to check the scaling relations
\beq
 \tilde\Pi_c(\eta,\eta',K)=K\hat\Pi(p,p')\,,\quad \tilde I_c(\eta,\eta',K)=K\hat I(p,p')
\eeq
where $p=-K\eta$ and $p'=-K\eta'$, which can be used to convert the above equations to the $p$-representation. One gets
\beq
\label{eq:Np1}
 \hat\Pi(p,p')=-\frac{\lambda}{6}\,(pp')^{d-3\over2}\!\!\int_\bq\frac{\hat G\left(qp,qp'\right)}{q}\frac{\hat G\left(rp,rp'\right)}{r},
\eeq
where $r=|\be+\bq|$, and the function $\hat I$ satisfies the integral equation
\beq
\label{eq:Np2}
 \hat I(p,p')=\hat\Pi(p,p')-i\int_{\hat\C} ds \,\hat\Pi(p,s )\hat I(s ,p').
\eeq
Finally the nonlocal part of the NLO self-energy reads
\beq
\label{eq:Np3}
 \hat\Sigma^{\rm NLO}_{\rm nl}(p,p')=\frac{\lambda}{3N}\,(pp')^{d-3\over2}\!\!\int_\bq\,\frac{r}{q}\,\hat G\left(qp,qp'\right)\hat I\left(rp,rp'\right)\!.
\eeq
Again one can check that Eqs. \eqn{eq:Np1} \eqn{eq:Np3} can be obtained by direct application of the diagrammatic rules in the $p$-representation as described in the previous section.

Let us finally write the above equation in terms of the explicit components on the momentum contour $\hat\C$. The product $\hat G^2$ in the expression \eqn{eq:Np1} of the function $\hat\Pi$ gives rise to the combinations $\hat F\hat F-{1\over4}\hat\rho\hat\rho$ for $\Pi_F$ and $2\hat F\hat\rho$ for $\Pi_\rho$. Similarly, the product $\hat G\hat I$ in \eqn{eq:Np3} gives $\hat F\hat I_F-{1\over4}\hat\rho\hat I_\rho$ for $\hat\Sigma^{\rm NLO}_F$ and $\hat F\hat I_\rho+\hat\rho\hat I_F$ for $\hat\Sigma^{\rm NLO}_\rho$. Finally, the contour integrals in \eqn{eq:Np2} are obtained from \Eqn{eq:exo2} as 
\bea
 \hat I_F(p,p')&=&\hat\Pi_F(p,p')-\!\int_{p'}^\infty \!\!ds \,\hat\Pi_F(p,s )\hat I_\rho(s ,p')\nn
 &+&\!\int_{p}^\infty \!\!ds \,\hat\Pi_\rho(p,s )\hat I_F(s ,p'),\\
 \hat I_\rho(p,p')&=&\hat\Pi_\rho(p,p')+\int^{p'}_{p} ds \,\hat\Pi_\rho(p,s )\hat I_\rho(s ,p').
\eea

We end this section by mentioning that the infinite series of bubble diagrams discussed here is actually related to the four-point vertex function in the large-$N$ limit. The latter is studied in Ref. \cite{4pt}, where the above integral equations are solved exactly in the limit of IR momenta $p,p'\ll1$, making extensive use of the $p$-representation.

\section{2PI approximation schemes}
\label{sec:2PI}

An important class of approximation schemes is based on 2PI functional methods \cite{Luttinger:1960ua,Baym:1962sx,deDominicis:1964zz,Cornwall:1974vz}. These provide systematic infinite resummations of selective sets of perturbative contributions and have proven a very useful tool in recent years to resum infrared divergences of bosonic theories in flat space-time at very high temperatures \cite{Blaizot:2003tw} or secular divergences of nonequilibrium field theory \cite{Berges:2004vw}. It has been shown that these methods are also useful in dealing with infrared and secular issues in de Sitter geometry \cite{Ramsey:1997qc,Riotto:2008mv,Garbrecht:2011gu,Serreau:2011fu}. We briefly recall the main ingredient of the 2PI formalism in a nonequilibrium setup \cite{Calzetta:1986cq,Berges:2004vw} and show how it can be formulated in the $p$-representation. 

2PI self-consistent approximation schemes are based on truncations or systematic expansions of the 2PI effective action, $\Gamma[\phi,G]$, a functional of both the one- and the two-point correlation functions of the theory in the quantum state under consideration $\phi_a(x)=\langle\varphi_a(x)\rangle$ and $G_{ab}(x,x')=\langle T_\C\varphi_a(x)\varphi_b(x')\rangle$. It can be parametrized as
\beq
\label{eq:2PIparam}
 \Gamma[\phi,G]=S[\phi]+{i\over2}{\rm Tr}{\rm Ln} G^{-1}+{i\over2}{\rm Tr}G_0^{-1}G+\Gamma_{\rm int}[\phi,G],
\eeq
where both the trace ${\rm Tr}$ and the logarithm ${\rm Ln}$ are to be understood in the functional sense. Here $S$ is the classical action, $iG_0^{-1}$ is the inverse free covariant propagator and $\Gamma_{\rm int}$ can be represented as the infinite sum of closed 2PI diagrams with lines $G$ and  vertices---including two-leg vertices---given by the shifted action $S[\phi+\varphi]$. Such diagrams with lines given by the exact propagator of the theory instead of the perturbative one are called skeleton diagrams. 

All vertex and correlation functions of the theory can be obtained from functional derivatives of the 2PI effective action evaluated at the solution of the equations of motion for both $\phi$ and $G$:
\beq
 \frac{\delta_c\Gamma[\phi,G]}{\delta\phi_a(x)}=0\,,\quad\frac{\delta_c\Gamma[\phi,G]}{\delta G_{ab}(x,x')}=0,
\eeq
where we define the covariant functional derivatives \cite{Ramsey:1997qc}
\bea
 {\delta_c\over\delta\phi_a(x)}&\equiv&{1\over\sqrt{-g(x)}}{\delta\over\delta\phi_a(x)},\\
 {\delta_c\over\delta G_{ab}(x,x')}&\equiv&{1\over\sqrt{-g(x)}}{1\over\sqrt{-g(x')}}{\delta\over\delta G_{ab}(x,x')}.
\eea
In particular, using the parametrization \eqn{eq:2PIparam}, the second equation gives the SD equation
\beq
\label{eq:SD2PI}
 G^{-1}=G_0^{-1}-\Sigma
\eeq
with 
\beq
 \Sigma_{ab}(x,x')=2i\frac{\delta_c\Gamma_{\rm int}[\phi,G]}{\delta G_{ba}(x',x)}.
\eeq
The point here is that the self-energy thus obtained is typically a nonlinear functional of the full propagator and one has to solve \Eqn{eq:SD2PI} self-consistently. This is where the nonperturbative nature of this approximation scheme enters. 

We now observe that 2PI approximation schemes are formulated in terms of two-point functions and skeleton diagrams. Thus, since the full propagator has the correct scaling \eqn{eq:prep}, all the considerations of previous sections concerning the SD equations and the diagrammatic rules in the $p$-representation hold. The only modification is that the free propagator is replaced by the full one, to be determined self-consistently by solving the SD equations.

For simplicity, let us consider $O(N)$ symmetric states, for which $\phi_a=0$, $G_{ab}=\delta_{ab} G$ and $\Sigma_{ab}=\delta_{ab}\Sigma$. In that case, the self-energy $\Sigma$ can be obtained from the functional $\Gamma_{\rm int}$ evaluated in the symmetric configuration $\phi_a=0$, $G_{ab}=\delta_{ab} G$ as 
\beq
 \Sigma(x,x')={2i\over N}\frac{\delta_c\Gamma_{\rm int}[\phi=0,G]}{\delta G(x',x)}.
\eeq
For instance a 2PI loop expansion at two-loop order gives
\beq
 \Gamma_{\rm int}[\phi=0,G]=-\frac{gN}{4}\int_xG^2(x,x)-i\frac{g'N}{8}\int_{xy}G^4(x,y),
\eeq
with $g=\lambda(N+2)/6N$ and $g'=-\lambda^2(N+2)/18N^2$ defined previously. One obtains for the self-consistent self-energy
\beq
 \Sigma(x,x')=-igG(x,x)\delta^{(D)}(x,x')+g'G^3(x,x')
\eeq
These expressions have the same structure as the standard 1PI one-loop expressions described in the previous section\footnote{In the 2PI loop-expansion, the diagram of Fig. \ref{fig:twolooplocal} is absent because of the 2PI character of the diagrammatic expansion, which in fact avoids possible double-counting. The missing two-loop contribution is now included in the self-consistent propagator.} and can thus be easily formulated in the $p$-representation. It is sufficient to replace free propagators by full ones in the diagrammatic rules.

Similarly, the 2PI $1/N$ expansion \cite{Aarts:2002dj} at NLO gives, up to an unphysical constant,
\beq
 \Gamma_{\rm int}[\phi=0,G]=-\frac{\lambda N}{4!}\int_xG^2(x,x)+{i\over2}{\rm Tr}{\rm Ln}D^{-1},
\eeq
where
\beq
 iD^{-1}(x,x')={3N\over \lambda}\left[\delta^{(D)}(x,x')-i\Pi(x,x')\right],
\eeq
with
\beq
 \Pi(x,x')=-{\lambda\over6}G^2(x,x').
\eeq
 
One easily checks that
\beq
 iD(x,x')=-{\lambda\over3N}\left[\delta^{(D)}(x,x')+iI(x,x')\right],
\eeq
where
\beq
 {I}(x,x')=\Pi(x,x')+i\int_z\Pi(x,z){I}(z,x').
\eeq
The corresponding self-consistent self-energy is given by
\beq
 \Sigma(x,x')=-i\sigma_0\delta^{(D)}(x,x')+{\lambda\over3N}G(x,x')I(x,x'),
\eeq
with
\beq
 \sigma_0={\lambda\over6}\left(1+{2\over N}\right)G(x,x).
\eeq
We see again that the structure of the equations is the same as those discussed in the previous section, the only change being that the LO propagator is now replaced by the full one. The $p$-representation of the 2PI $1/N$ expansion is thus obviously obtained. 

\section{Conclusions}

We have developed a systematic method for computing correlation and vertex functions of a scalar field in de Sitter space. It  exploits both the simplifications due to de Sitter symmetries and the power of a momentum representation, e.g., for writing spatial convolution integrals as simple products. The method relies on the particular way momentum redshift is encoded in de Sitter correlators, see \Eqn{eq:rigid}, which implies a one-to-one correspondence between time and physical momentum and allows one to trade one for the other.

This method is particularly well adapted to describing two-point functions---for which it reduces the number of independent variables from $3$ to $2$ as compared to the comoving momentum representation---and thus to all approximation schemes based on the use of the latter. This includes standard expansion schemes in QFT such as the loop, or the $1/N$ expansions, but also resummed approximation schemes based on 2PI techniques. For what concerns two-point correlators, our approach effectively reduces the problem to a ($0+1$)-dimensional one, where physical momentum plays the role of the ``time'' variable. 

We emphasize that the resulting equations are well suited to analytical approximations as well as to numerical implementation. This is particularly important for studies of infrared/secular issues in de Sitter, which require infinite resummations and thus possibly numerical work, or for studies of trans-Planckian issues which may require nonperturbative calculations of unequal time (unequal momentum) correlators, in the same spirit as the calculation of damping and thermalization effects from 2PI techniques in flat space-time \cite{Berges:2004vw}. 

Work in these directions has been pursued and the results will soon be presented \cite{4pt,Gautier}. In \cite{4pt}, we studied the four-point vertex function of an $O(N)$ scalar field. In the large-$N$ limit, 
this vertex is given by an infinite series of bubble diagrams, each of which exhibits large IR logarithms, typical of de Sitter space. Exploiting the method presented here, we found by analytical means that the resulting resummation of IR logarithms leads to a modified power law in the deep IR, analogous to the generation of an anomalous dimension in critical phenomena.

Finally, we mention that the present method, since it does not exploit all de Sitter symmetries, allows one to treat deformations of de Sitter which are compatible with the $p$-representation. In fact the usefulness of
this representation was first understood in the context of theories where the Lorentz group is violated
by dispersive or dissipative effects occurring in the UV sector \cite{Busch:2012ne,ABP}. Interesting extensions of the present work also include the discussion of the $p$-representation for fields of higher spin and/or theories with derivative couplings.

\section*{Acknowledgements}

We thank Xavier Busch and Florian Gautier for their interesting and useful remarks.

\appendix

\section{Time and momentum contours}
\label{appsec:contours}

The notion of a closed time contour \cite{Schwinger:1960qe,Bakshi:1962dv,Keldysh:1964ud,Chou:1984es} provides a convenient way to deal with the various components of $n$-point functions in a general nonequilibrium setup. Alternative formulations employ tensor structures, such as the so-called $\pm$ or Keldysh basis \cite{Keldysh:1964ud,Chou:1984es}. 

The closed contour $\C=\C^+\cup\C^-$ in conformal time can be described by a mapping $\eta(s)$ of the real axis $\mathbb{R}$ onto the contour $\C$, such that the negative and positive real axis $\mathbb{R}^\mp$ are mapped onto the upper and lower branches $\C^\pm$ respectively:
\beq
 \eta(s)\in\C^\pm\quad{\rm for}\quad s\in\mathbb{R}^\mp,
\eeq
with $\eta(s)$ a monotonously increasing (decreasing) function on the negative (positive) real axis. The ordering along $\C$ is defined as an ordering along $\mathbb{R}$:
\beq
 {\rm sign}_\C(\eta-\eta')\equiv{\rm sign}_\C(\eta(s)-\eta(s'))={\rm sign}(s-s').
\eeq
The integral of a function $f(\eta)\equiv f(\eta(s))$ along the contour is defined as
\beq
 \int_\C d\eta f(\eta)=\int_\mathbb{R}ds\,\frac{d\eta}{ds}\,f(\eta(s)).
\eeq
We define the Dirac delta function on $\C$ as
\beq
 \delta_\C(\eta-\eta')\equiv\delta_\C(\eta(s)-\eta(s'))=\left(\frac{d\eta}{ds}\right)^{\!\!-1}\delta(s-s')
\eeq
in such a way that
\beq
 \int_\C d\eta'\delta_\C(\eta-\eta')f(\eta')=\int_\mathbb{R}ds\,\delta(s-s')\,f(\eta(s'))= f(\eta).
\eeq
Notice the properties
\beq
 \frac{d}{d\eta}{\rm sign}_\C(\eta-\eta')=2\delta_\C(\eta-\eta')
\eeq
and
\beq
 \delta_\C(\eta-\eta')=\pm\delta(\eta-\eta')\quad{\rm if}\quad \eta,\eta'\in\C^\pm.
\eeq

The contour $\hat\C$ in momentum is defined in an analogous way. We introduce the function ($K$ is a constant)
\beq
 p(s)=-K\eta(s),
\eeq
which maps the negative and positive real axis $\mathbb{R}^\mp$ onto the upper and lower branches $\hat\C^\pm$ respectively:
\beq
 p(s)\in\hat\C^\pm\quad{\rm for}\quad s\in\mathbb{R}^\mp.
\eeq
Here, $p(s)$ is a monotonously decreasing (increasing) function on the negative (positive) real axis. The ordering along $\hat\C$, defined as
\beq
 {\rm sign}_{\hat\C}(p-p')\equiv{\rm sign}_{\hat\C}(p(s)-p(s'))={\rm sign}(s-s'),
\eeq
is such that
\beq
 {\rm sign}_{\hat\C}(p-p')={\rm sign}_{\C}(\eta-\eta').
\eeq
For a given function
\beq
 f(\eta(s))=\hat f(p(s))
\eeq
we define the integral
\beq
 \int_{\hat\C} dp\hat f(p)=\int_\mathbb{R}ds\,\frac{dp}{ds}\,\hat f(p(s))=-K\int_\C d\eta f(\eta).
\eeq
We define a Dirac delta function on $\hat\C$ as before:
\beq
 \delta_{\hat\C}(p-p')\equiv\delta_{\hat\C}(p(s)-p(s'))=\left(\frac{dp}{ds}\right)^{\!\!-1}\delta(s-s'),
\eeq
such that
\beq
 \int_{\hat \C} dp'\delta_{\hat\C}(p-p')\hat f(p')=\int_\mathbb{R}ds\,\delta(s-s')\,\hat f(p(s'))= \hat f(p).
\eeq
One has the properties
\beq
 \frac{d}{dp}{\rm sign}_{\hat\C}(p-p')=2\delta_{\hat\C}(p-p')
\eeq
and
\beq
 \delta_{\hat\C}(p-p')=\mp\delta(p-p')\quad{\rm if}\quad p,p'\in\hat\C^\pm.
\eeq
Finally, notice the relation
\beq
 \delta_\C(\eta-\eta')=-K\delta_{\hat\C}(p-p').
\eeq

\section{Higher order correlators and vertex functions}
\label{appsec:higher}

In this appendix, we discuss the consequences of the $p$-representation \eqn{eq:prep} of two-point functions for higher order correlation and vertex functions, as obtained from perturbative diagrams. The connected time-ordered $n$-point correlator is
\beq
 G^{(n)}(x_1,\ldots,x_n)=\langle T_\C\varphi(x_1)\cdots\varphi(x_n)\rangle_c
\eeq
and we define the covariant $n$-point proper vertex functions $\Gamma^{(n)}$ from the field expansion of the 1PI effective action
\beq
 \Gamma[\varphi]=\sum_{n\ge0} \int_{x_1,\ldots,x_n}\varphi(x_1)\cdots\varphi(x_n)\Gamma^{(n)}(x_1,\ldots,x_n).
\eeq
Equivalently
\beq
 \left.\Gamma^{(n)}(x_1,\ldots,x_n)=\frac{\delta_c^n\Gamma[\varphi]}{\delta\varphi(x_1)\cdots\delta\varphi(x_n)}\right|_{\varphi=\phi}
\eeq
with the covariant functional derivative \eqn{eq:covfuncder}. The corresponding conformally rescaled quantities are given by
\beq
 G^{(n)}(x_1,\ldots,x_n)=[a(\eta_1)\ldots a(\eta_n)]^{-{d-1\over2}}G^{(n)}_c(x_1,\ldots,x_n)
\eeq
and
\beq
 \Gamma^{(n)}(x_1,\ldots,x_n)=[a(\eta_1)\ldots a(\eta_n)]^{-{d+3\over2}}\Gamma^{(n)}_c(x_1,\ldots,x_n).
\eeq
The comoving momentum representation is defined as
\beq	
 G^{(n)}_c(x_i)=\int_{\bK_1,\ldots,\bK_n}e^{i\sum_{j=1}^n\bK_j\cdot\bX_j}\bar G_c^{(n)}(\eta_i,\bK_i),
\eeq
and similarly for $\Gamma^{(n)}_c$, where we denote collectively $x_i\equiv x_1,\ldots,x_n$ and similarly for $\eta$'s and $\bK$'s.
Spatial homogeneity in comoving coordinates implies that
\beq
  \bar G_c^{(n)}(\eta_i,\bK_i)=(2\pi)^d\delta^{(d)}\left(\sum_{i=1}^n\bK_i\right)\tilde G_c^{(n)}(\eta_i,K,\hat \bK_i).
\eeq
and similarly for $\Gamma^{(n)}_c$. Spatial isotropy further reduces the number of independent momentum variables.
Here, we extracted an explicit arbitrary comoving momentum scale $K$ from the set of variables $\bK_i$ and define $\hat \bK_i=\bK_i/K$. For instance, one can choose to single out one momentum: $K=|\bK_1|$, or take the symmetric combination $K=\sqrt{\bK_1^2+\ldots+\bK_n^2}$, etc.

Now, we consider a given diagram contributing to $\Gamma^{(n)}_c$ with $V$ internal vertices. We apply the diagrammatic rules recalled in Sec. \ref{sec:diag} and perform the same rescaling of all---internal and external---variables as in Sec. \ref{sec:diag} using the scale $K$ defined above. Since there are now $n$ vertices attached to the endpoints $x_1,\ldots,x_n$, there are only $V-n$ time integrals over internal vertices. We thus get the overall factor
\bea
 &&\frac{K^{d(V-n)}}{K^{d(V-1)}}\times[(-\eta_1)\cdots(-\eta_n)]^{-{d+3\over2}}\nn
 \label{appeq:facn}
 &&\quad\qquad=K^{d+\frac{n(3-d)}{2}}\times (\hat p_1\cdots \hat p_n)^{-{d+3\over2}},
\eea
which generalizes \Eqn{eq:factorover} for $n\neq2$. Here, we introduced the variables $\hat p_i=-K\eta_i=p_i/|\hat\bK_i|$, with $p_i=-|\bK_i|\eta_i$ the physical external momenta. The factor \eqn{appeq:facn} is independent of $V$ and thus of the diagram one considers. We conclude that the $n$-point vertex function admits the scaling (or $p$-representation)
\beq
\label{appeq:prepGamma}
 \tilde\Gamma_c^{(n)}(\eta_i,K,\hat \bK_i)=K^{d+\frac{n(3-d)}{2}}\hat\Gamma^{(n)}(\hat p_i,\hat \bK_i).
\eeq
A similar analysis for the $n$-point correlator $\tilde G_c^{(n)}$ leads to the $p$-representation 
\beq
\label{appeq:prepG}
 \tilde G_c^{(n)}(\eta_i,K,\hat \bK_i)=K^{d-\frac{n(d+1)}{2}}\hat G^{(n)}(\hat p_i,\hat \bK_i).
\eeq
Eqs. \eqn{appeq:prepGamma} and \eqn{appeq:prepG} generalize Eqs. \eqn{eq:selfscale} and \eqn{eq:prep} for two-point functions (for which a natural choice for the scale $K$ is $K=|\bK_1|=|\bK_2|$).

\section{The auxiliary field formalism}
\label{appeq:NLOaux}

The $1/N$ expansion discussed in Sec. \ref{sec:N} can be conveniently formulated by introducing an auxiliary field $\chi$ corresponding to the composite operator $\varphi^2$ \cite{Coleman:1974jh,Root:1974zr,Cooper:1994hr,Aarts:2002dj,Cooper:2004rs}. Consider the action
\beq
\label{appeq:chi}
 S[\varphi,\chi]={1\over2}\int_x\left\{\varphi_a\left(\square-m_{\rm dS}^2\right)\varphi_a+{3N\over\lambda}\chi^2-\chi\varphi_a\varphi_a\right\}.
\eeq
The equation of motion for the auxiliary field $\chi$ is a constraint equation: $\chi=\lambda\varphi_a\varphi_a/6N$. Integrating out the field $\chi$, one recovers the original action \eqn{eq:classical}. The field $\chi$ is a convenient bookkeeping device to count factors of $N$ and thus to organize the $1/N$ expansion.

\begin{figure}[h!]  
\epsfig{file=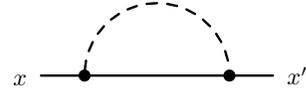,width=4cm}
 \caption{\label{fig:nlo2} The NLO contribution to the self-energy $\Sigma(x,x')$ in the $1/N$ expansion in the auxiliary field formulation. The plain line represents the propagator $G$ of the field $\varphi$ whereas the dashed line denotes the propagator $D$ of the auxiliary composite field $\chi\propto\varphi^2$. The latter resums the infinite series of bubble diagrams depicted in Fig. \ref{fig:bubbles}; see \Eqn{appeq:DpropI}. The dot represents the $\chi\varphi^2$ vertex.}
\end{figure}

\begin{figure}[h!]  
\epsfig{file=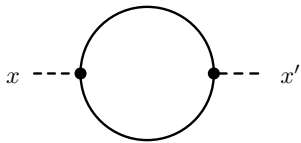,width=4cm}
 \caption{\label{fig:nlo3} The auxiliary field self-energy at LO in the $1/N$ expansion.}
\end{figure}

In this formulation, the NLO contribution to the self-energy is given by the single one-loop diagram of Fig. \ref{fig:nlo2}, corresponding to the expression:
\beq
\label{appeq:NLObis}
 \Sigma^{\rm NLO}(x,x')=-G(x,x')D(x,x')
\eeq
where $G$ is the propagator of the field $\varphi$ at LO and
\beq
 D(x,x')=\langle T_\C\chi(x)\chi(x')\rangle
\eeq
is the propagator of the composite field $\chi$. At this order of approximation, the $\chi$ self-energy is given by the diagram of Fig. \ref{fig:nlo3}, which corresponds to $-{N\over2}G^2(x,x')={3N\over\lambda}\Pi(x,x')$, in terms of the one-loop bubble function $\Pi(x,x')$; see \Eqn{eq:Pifunc}. One thus has
\beq
 iD^{-1}(x,x')={3N\over\lambda}\left[\delta^{(D)}(x,x')-i\Pi(x,x')\right],
\eeq
where the first term on the right-hand side is the tree-level contribution; see \Eqn{appeq:chi}.
Using \Eqn{eq:Ifunc}, it is easy to check that the propagator $D(x,x')$ is thus related to the infinite series of bubbles $I(x,x')$ as
\beq
\label{appeq:DpropI0}
 D(x,x')={i\lambda\over3N}\left[\delta^{(D)}(x,x')+iI(x,x')\right].
\eeq
Inserting this expression in \Eqn{appeq:NLObis}, one recovers the result \eqn{eq:NLO1}-\eqn{eq:NLOnl}.

Introducing the conformally rescaled correlator as 
\beq
 D(x,x')=\left[a(\eta)a(\eta')\right]^{-{d+1\over2}}D_c(\eta,\eta',|\bX-\bX'|),
\eeq
one gets, in comoving momentum space,
\beq
\label{appeq:DpropI1}
 \tilde D_c(\eta,\eta',K)={i\lambda\over3N}\left[\delta_\C(\eta-\eta')+i\tilde I_c(\eta,\eta',K)\right].
\eeq
The $p$-representation of the auxiliary field correlator is obtained as
\beq
 \tilde D_c(\eta,\eta',K)=K\hat D(p,p')
\eeq
with $p=-K\eta$ and $p'=-K\eta'$. One has, in particular,
\beq
\label{appeq:DpropI}
 \hat D(p,p')={i\lambda\over3N}\left[-\delta_{\hat\C}(p-p')+i\hat I(p,p')\right].
\eeq

Finally, we mention that the 2PI $1/N$ expansion, discussed above, can also be formulated in the auxiliary field formalism  \cite{Aarts:2002dj,Cooper:2004rs}. The relevant equations are formally the same as presented here, with the understanding that the correlators $G$ and $D$ are the NLO ones---instead of the LO ones---to be determined self-consistently from the SD equations at NLO.

\end{document}